\newcommand{\be}{\begin{equation}}
        \newcommand{\ee}{\end{equation}}
        \newcommand{\ba}{\begin{eqnarray}}
        \newcommand{\ea}{\end{eqnarray}}
        \newcommand{\ban}{\begin{eqnarray*}}
        \newcommand{\ean}{\end{eqnarray*}}
         
\newcommand{\R}{{\mathbb R}}  
\newcommand{\C}{{\mathbb C}}

\newtheorem{definition}{Definition} 
\newtheorem{theorem}{Theorem} 
\newtheorem{lemma}{Lemma} 
\newtheorem{corollary}{Corollary}

\documentclass[12pt]{article}
\usepackage {graphicx,latexsym}
\usepackage{amsfonts}

\begin{document}

\title{Geometry of $C$-flat connections, coarse graining 
and the continuum limit}

\author{Jorge Mart\'\i{}nez, 
Claudio Meneses and
Jos\'e A. Zapata
\footnote{\ttfamily yorch@matmor.unam.mx, 
claudio@matmor.unam.mx, 
zapata@math.unam.mx
}
\\  %%for iop this line is commented%%
%\address{ 
Instituto de Matem\'aticas UNAM \\ 
A.P. 61-3, Morelia Mich. 58090, M\'exico
}
\date{}
\maketitle
\abstract{
A notion of effective gauge fields which does not involve a background metric is introduced. 
The role of scale is played by cellular decompositions of the base manifold. Once a cellular decomposition is chosen, the corresponding space of effective gauge fields is the space of 
flat connections with singularities on its codimension two skeleton, 
${\cal A}_{C\!\hbox{\footnotesize -flat}} / \bar{\cal G}_{M,\star} \subset 
\bar{\cal A}_M / \bar{\cal G}_{M,\star}$.  
If cellular decomposition $C_2$ is finer than cellular decomposition $C_1$, 
there is a coarse graining map $\pi_{C_2 \to C_1}
: {\cal A}_{C_2\!\hbox{\footnotesize -flat}}/ \bar{\cal G}_{M,\star} \to 
{\cal A}_{C_1\!\hbox{\footnotesize -flat}} / \bar{\cal G}_{M,\star}$. 
We prove that the triple 
$({\cal A}_{C_2\!\hbox{\footnotesize -flat}}/ \bar{\cal G}_{M,\star} , 
\pi_{C_2 \to C_1} , 
{\cal A}_{C_1\!\hbox{\footnotesize -flat}} / \bar{\cal G}_{M,\star})$ 
is a principal fiber bundle with a preferred global section given by the natural inclusion map 
$i_{C_1 \to C_2}: 
{\cal A}_{C_1\!\hbox{\footnotesize -flat}} / \bar{\cal G}_{M,\star} \to 
{\cal A}_{C_2\!\hbox{\footnotesize -flat}}/ \bar{\cal G}_{M,\star}$. 

Since the spaces ${\cal A}_{C\!\hbox{\footnotesize -flat}} / \bar{\cal G}_{M,\star}$
are partially ordered (by inclusion) and this order is directed in the direction of refinement, we can define a continuum limit, 
$C \to M$. We prove that, in an appropriate sense, 
$\lim_{C \to M} {\cal A}_{C\!\hbox{\footnotesize -flat}} / \bar{\cal G}_{M,\star} = \bar{\cal A}_M / \bar{\cal G}_{M,\star}$. 
We also define a construction of measures in 
$\bar{\cal A}_M / \bar{\cal G}_{M,\star}$ as the continuum limit (not a projective limit) 
of effective measures. 
}

\section{Motivation}
The Wilsonian renormalization group is fundamental in standard constructions of 
quantum field theories. 
The notions of {\em effective theories at given scales, 
coarse graining and the ultraviolet/continuum limit} 
are essential for that scheme. 

There are important physical systems for which there is no natural concept of scale which can be used to describe 
an effective theory as needed by standard renormalization group ideas. 
Any system that includes gravitational 
phenomena is in this category of systems because a scale is defined 
through a metric which in these cases is a dynamical variable. Thus, a definition of
effective theories and coarse graining calls for%
\footnote{
It is of course possible to introduce a fiducial metric and use its induced notion of scale to define effective theories and coarse graining. In fact this alternative has been developed in recent years \cite{fiducialMetric}.
} 
an ``extension of the concept of scale."

We will study a novel proposal of effective theories and coarse graining for gauge theories which emulates structures from lattice gauge theory 
to situations that are free of a background metric.  
In standard lattice gauge theory one works with a family of 
effective theories (labeled by increasingly finer lattices) describing the same system. 
The key ingredient that fine tunes all these theories with observation and with each other is a renormalization procedure. 

General relativity can be formulated as a gauge theory, but if one wishes to represent the diffeomorphism symmetry it is impossible to use a single embedded lattice to host the theory. A solution to this problem is given by the kinematics of loop quantization ``which is a lattice gauge theory for a lattice that is infinitely refined" (for a precise statement see \cite{Proposal}). 
Since there was a single lattice -- even if infinitely refined -- 
there was no 
sequence of effective theories 
that let one implement Wilson's renormalization group and that in a continuum 
limit selected a dynamics for the loop quantized theory.

A goal of our research program \cite{Proposal} is to provide a 
family of effective theories connected to each other by coarse graining maps 
that average away fluctuations 
and whose infinite refinement limit takes us to a loop quantized theory. 
The dynamics of the continuum limit will be defined only after  fine tuning all the effective theories by a  renormalization procedure. 
The proposal is simple and natural in the context of loop quantization. 
However, there is no claim of uniqueness; we know other families of effective theories with similar structure. 
Physical applications are still being developed. The most illustrative result at the moment is that when the same ideas are applied to the Ising model one recovers the standard renormalization by blocking. In addition, 
an extension to irregular lattices of the 
the blocking and bond-moving procedure of Migdal and Kadanoff is natural in our framework \cite{Ising}. 

A regularization procedure is included in our framework 
because the configuration spaces of the effective theories lie inside the space of generalized connections, ${\cal A}_{\rm ``scale"} \subset \bar{\cal A}_M$. Thus any observable in the continuum (cylindrical function) is automatically regularized to an observable of the effective theory. This same feature lets us define the continuum limit of the effective theories. For example, the vacuum expectation value of a cylindrical function in the continuum $< f>_M$ is defined as the continuum limit of the vacuum expectation value of the same function evaluated in the effective theory defined at a given ``scale", 
``$< f>_M= \lim_{{\rm ``scale"} \to 0}< f|_{{\cal A}_{\rm ``scale"}}>_{\rm ``scale"}$."
Our structure should be compared with others using the restriction of the space of connections to a fixed graph (embedded lattice) ${\cal A}_{\gamma}$ as the home for an effective theory. In that case one would have to provide a separate regularization of observables from $\bar{\cal A}_M$ to act on each of the spaces ${\cal A}_{\gamma}$ contained in the sequence used to define the continuum limit. 
More over, these family of regularization procedures would have to satisfy compatibility conditions to define a continuum theory. 

The specific goal of this paper is to present a geometrical study of a family of  configuration spaces of effective theories and of the coarse graining maps that relate them to each other and to the space of generalized connections of loop quantization. 
The aim is to provide a solid ground for implementing a Wilsonian renormalization 
in a framework that is independent of a background metric.  
Eventually,  this will serve to define the dynamics (physical measure) of loop quantized theories as the continuum limit of the dynamics of our effective theories.

%Plan of the article. 
This article has the following organization. 
The next section introduces $C$-flat connections as effective gauge fields. It also 
studies the geometry of a $C$-flat connection and of the space of $C$-flat connections, as well defines the continuum limit 
$C \to M$. 
Section 3 defines coarse graining maps, studies the resulting geometry and 
writes an ``exact renormalization prescription" that links different effective measures. 
Section 4 
is about the action of diffeomorphisms on the spaces of $C$-flat connections; it also defines a space of diffeomorphism invariant effective configurations. 
Finally, the appendix 
contains detailed definitions of some preliminary material.

\section{A model of effective gauge fields}

%%%%%%%%Moved to first section 
%SCALE AND HOW MUCH WE MEASURE/HOW MUCH 
%REGULARITY WE ASSUME
The access to only finitely many measurement devises implies 
that only partial knowledge of the system is available. 
In order  to have 
a presumed state of the system {\em in the continuum} for each set of measurements, 
one complements this partial knowledge with regularity assumptions. 
A set of measuring devises and complementary regularity assumptions 
``turn on" some degrees of freedom at a given ``measuring scale." 
This intuitive idea of  ``measuring scale" is behind all our definitions.

%ANTICIPATION OF GEOM. DESCRIPTION OF C-CONFIGURATIONS
In this paper, the role of scale is played by cellular decompositions of the base manifold. Once a cellular decomposition is chosen, there are regularity assumptions tailored to it. The idea is that inside each cell of the cellular decomposition the connection will be as regular as possible. 
Ironically these regularity assumptions take us to 
distributional configurations: configurations that are not smooth connections 
but generalized connections. The particular kind of generalized connections that we work with are flat everywhere except for the codimension two cells of the cellular decomposition. The available observables at this ``measuring scale" are holonomies along the links of a lattice constructed from the cellular decomposition.

%DEF OF C-FLAT CONN. 
\begin{definition}[$C$-flat connections]
A generalized connection is considered to be $C$-flat, 
$A \in {\cal A}_{C\!\hbox{\footnotesize -flat}} \subset \bar{\cal A}_M$,
if and only if its restriction to 
each cell of $C$ is flat. 
\end{definition}
Notice first that the cells of a cellular decomposition are disjoint, 
$c_{\alpha} \cup  c_{\beta} = \emptyset \hbox{ if } \alpha \neq \beta$. 
Thus, there are many non flat connections which are $C$-flat. 

Also notice that in the context of generalized connections the concept of flatness is phrased in terms of the path independence of parallel transport. With this in mind our definition may be clear as it stands, but the following characterization in terms of the holonomy maps induced by $C$-flat connections is most useful in the rest of the paper. 

The presence of a cellular decomposition $C$ induces a natural equivalence relation among directed paths. Two oriented paths $\gamma_1, \gamma_2$ are defined to be 
{\em $C$-equivalent}, $\gamma_1 \sim_C \gamma_2$, if the two sequences of cells (of any dimension) 
induced by traversing the curves coincide. 
\begin{lemma}\label{holonomiesAndCequiv}
${\cal A}_{C\!\hbox{\footnotesize -flat}} / \bar{\cal G}_{M, \star}$ 
can be characterized as 
the subset of generalized connections modulo gauge transformations, 
$\bar{\cal A}_M / \bar{\cal G}_{M, \star}$, 
whose elements are those and only those which satisfy 
\[
[A](\gamma_1) = [A](\gamma_2)
\]
for any 
two closed oriented paths based at $\star \in M$ 
which are $C$-equivalent 
\[
\gamma_1 \sim_C \gamma_2 .
\]
\end{lemma}
The proof of this lemma is a simple application of the reconstruction theorems 
\cite{reconstruction} to our context.

In the rest of this section we will talk about three different subjects in corresponding subsections. In the first subsection we will describe the geometry of single flat connections; in the second we will elaborate on the space 
${\cal A}_{C\!\hbox{\footnotesize -flat}}$, and in the third we will present configuration spaces that are relevant for the continuum limit in our framework.

\subsection{Geometry of a $C$-flat connection}
%REMARK ON DISTRIBUTIONALITY OF C-FLAT CON AND GEO DESCR
A good starting point for studying the geometry of $C$-flat connections is the previous lemma; by using it, it is easy to see that 
the ``regularity" assumptions yield generalized connections that may not even be continuous. 
Consider, for example, a smooth deformation of a given closed curve. According to our lemma, once a $C$-flat connection is chosen, 
the holonomy of the curve would be independent of the deformation unless its $C$-equivalence class changes. At that point of the deformation process the holonomy may experience a drastic change. 

Moreover, since $\bar{\cal G}_{M,\star} ({\cal A}_{C\!\hbox{\footnotesize -flat}}) = 
{\cal A}_{C\!\hbox{\footnotesize -flat}}$ and $\bar{\cal G}_{M,\star}$ includes discontinuous gauge transformations, we see that even 
the gauge equivalence class of 
an everywhere flat connection has representatives which are not continuous. 

However, the distributional nature of our connections modulo gauge is of a very tame type. 
Essentially they are the familiar spaces of flat connections with ``conical" singularities along the codimension two skeleton of $C$. 
To see that this is the case consider a small loop contained in a single cell of maximum dimension and deform it continuously until it crosses one codimension one cell. The holonomy along the loop before the deformation was the identity and after the deformation it is still the identity because the loop is $C$-equivalent to another loop constructed as $l = c^{-1} \circ c$ (where $c$ is an open curve starting in one maximal dimension cell, crossing a codimension one cell, and ending at another maximal dimension cell). Then the codimension one cells cannot host curvature singularities. However, 
drastic changes in the holonomy during continuous deformations may result from crossing (or hitting) cells of codimension greater or equal to two%
\footnote{
In our framework also the holonomies along curves that pass trough singularities are defined. In some contexts these non generic curves may be regarded as not important. 
}. 
This observation is formally 
stated as a direct corollary of lemma \ref{holonomiesAndCequiv} and the reconstruction theorems \cite{reconstruction}. 
\begin{corollary}
For any $[A] \in {\cal A}_{C\!\hbox{\footnotesize -flat}}/\bar{\cal G}_{M, \star}$ there is a unique (up to fiber bundle equivalence) smooth bundle $\Phi$ over $M- C^{(n-2)}$ and a representative $A \in [A]$ 
such that 
\[
A|_{M- C^{(n-2)}} \in {\cal A}_{M- C^{(n-2)},\Phi}^{\infty} . 
\]
Moreover, in this domain its curvature is defined and vanishes. 
\end{corollary}

One can expect that the topological charges that characterize the topology and geometry of 
such singularities will play an important role when investigating 
the dynamical properties of certain types of fields. Within our framework it would be natural to choose stronger regularity assumptions to make the space of effective configurations at scale $C$ the space of $C$-flat connections with a restricted type of topological charges at the singularities.

\subsection{Geometry of the space of $C$-flat connections}
%STUDY OF ${\cal A}_{C\!\hbox{\footnotesize -flat}}$
%DIMENSION, DIM MOD GAUGE, equiiv to ${\cal A}_{L(C)} / {\cal G}_{L(C)}$ 
%TOPOLOGICAL PROPERTIES AS SUBSET OF GEN CONN. 
The space ${\cal A}_{C\!\hbox{\footnotesize -flat}}$ is infinite dimensional, but we can expect 
${\cal A}_{C\!\hbox{\footnotesize -flat}} / \bar{\cal G}_{M, \star}$ to be finite dimensional because flat connections modulo gauge do not have any local degrees of freedom. We will show that this is indeed the case by conveniently characterizing 
gauge equivalence classes of 
$C$-flat connections in terms of lattice gauge theory connections. 

The starting point is again
lemma \ref{holonomiesAndCequiv}. 
Observe that the lattice dual to $C$ contains representatives of 
the $C$-equivalence classes of 
``most" curves, but not all. Hence the space of connections on a lattice dual of $C$ 
has almost all the holonomy information in 
${\cal A}_{C\!\hbox{\footnotesize -flat}} / \bar{\cal G}_{M, \star}$, but some is missing. Then we complete the dual lattice to include curves $C$-equivalent to curves passing through 
cells of codimension bigger than one; call the resulting lattice $L(C)$. 
By definition oriented paths in $L(C)$ label $C$-equivalence classes of oriented paths in $M$. 
In the case of simplicial cellular decompositions $L(C)$ is the one skeleton of their barycentric subdivision, $L(C)={\rm Sd}(C)^{(1)}$ (see the appendix for a definition). In the following figure we present another simple example. 
\begin{figure}[!ht]
\centerline {
\includegraphics[width=7cm]{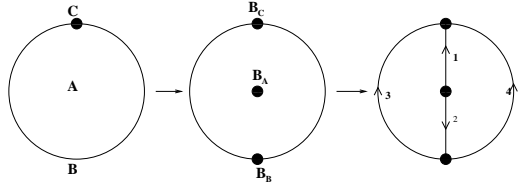}\hspace{1.5cm}
}
\caption{Construction of $L(C)$}\label{F:2}
\end{figure}

An embedding ${\rm Emb}_{L(C)}: L(C) \to M$ can be used to 
define a new fiber bundle with (one dimensional) base $L(C)$ 
starting from the fiber bundle $(E, \varphi, M)$. The new total space would be 
$\varphi^{-1}({\rm Emb}_{L(C)} [L(C)])$, 
and projection $\varphi |_{{\rm Emb}_{L(C)} [L(C)]}$. 
Apart from pulling back the bundle, 
${\rm Emb}_{L(C)}$ pulls back connections from $\bar{\cal A}_M$ to 
the space of connections on the new bundle, 
${\rm Emb}_{L(C)}^{\star}: \bar{\cal A}_M \to {\cal A}_{L(C)}$.

Clearly not all embeddings are of our interest. To maintain the meaning of directed paths in $L(C)$ as representatives of $C$-equivalence classes of directed paths in $M$ we restrict ourselves to {\em representative embeddings}. These are 
embeddings ${\rm Emb}_{L(C)}: L(C) \to M$ such that the inverse image of any oriented path in $M$ is their corresponding $C$-equivalence class. 

We denote the groups of based oriented loops on $M$ by ${\cal P}_{M, \star}$; we also denote the induced group of $C$-equivalent classes of loops by 
${\cal P}_{C, \star}$ and by ${\cal P}_{L(C), b}$ (${\rm Emb}_{L(C)}(b) = \star$) the analogous object in $L(C)$. 

Note that  
for any representative embedding 
\[
{\rm Emb}_{L(C)}: {\cal P}_{L(C), b} \to {\cal P}_{C, \star}
\]
is an isomorphism independent of the choice of representative embedding (with ${\rm Emb}_{L(C)}(b) = \star$). 
Thus we also have the natural isomorphism 
\[
{\rm Emb}_{L(C)}^{\star}: {\rm Hom}( {\cal P}_{C, \star}, G) \to 
{\rm Hom}( {\cal P}_{L(C), b}, G) , 
\]
where ${\rm Hom}( {\cal P}_{C, \star}, G)$ is naturally embedded in 
${\rm Hom}( {\cal P}_{M, \star}, G)$. 

Using the reconstruction theorems \cite{reconstruction} this can be stated in a more familiar language as follows: 
there is a natural isomorphism 
\[
{\rm Emb}_{L(C)}^{\star}:
{\cal A}_{C\!\hbox{\footnotesize -flat}} / \bar{\cal G}_{M, \star} 
\to {\cal A}_{L(C)} / {\cal G}_{L(C), b} 
\]
which is independent of the choice of representative embedding 
${\rm Emb}_{L(C)}: L(C) \to M$ (with ${\rm Emb}_{L(C)}(b) = \star$). 

It is important to remark that simplicity appears only at the gauge invariant level; 
the space ${\cal A}_{C\!\hbox{\footnotesize -flat}}$ is not isomorphic to a lattice gauge theory configuration space. 
Working on infinite dimensional quotient spaces is difficult, but we know that the complications must be inessential. This motivates us to study the same spaces in a partially gauge fixed context. 
It can be thought of as a strengthening of our previous regularity assumptions with extra gauge fixing conditions. 

%${\cal A}_{L(C)} in a local triviallization PARTIAL GAUGE FIX AND C-FLAT
Once a local trivialization is chosen, 
we can assign group elements to holonomy mappings of open paths. 
Using this fact, 
a partially gauge fixed version of the space of 
$C$-flat connections is easily characterized as follows: \\
we say that 
$A \in {\cal A}_{C\!\hbox{\footnotesize -flat},A_0}$ if and only if 
the generalized connection $A \in \bar{\cal A}_{M}$ is such that 
the holonomies along any (possibly open) oriented paths are equal, 
$A(\gamma_1) = A(\gamma_2) \in G$, whenever 
$\gamma_1 \sim_C \gamma_2$. 
The space ${\cal A}_{C\!\hbox{\footnotesize -flat},A_0}$ depends on 
the local trivialization. It is more convenient to say that it depends on an auxiliary flat connection $A_0$ (the one induced by the local trivialization). 
Using this notation it is clear that generic gauge transformations do not leave it invariant, but 
$g({\cal A}_{C\!\hbox{\footnotesize -flat},A_0}) = {\cal A}_{C\!\hbox{\footnotesize -flat},g(A_0)}$. The gauge transformations that do leave it invariant form a finite dimensional subgroup of the group of gauge transformations denoted by ${\cal G}_{C,A_0, \star} \subset \bar{\cal G}_{M, \star}$. We can also define this space of residual gauge transformations in terms of the local trivialization as 
composed by gauge transformations such that $g(p)= g(q) \in G$ whenever $p$ and $q$ are in the same cell. 

It is clear that, after the partial gauge fixation 
$P_{A_0}: {\cal A}_{L(C)} \to 
{\cal A}_{L(C), A_0}$, these spaces of connections can be parametrized by the assignment of group elements to $C$-equivalence classes of paths, or more conveniently to paths in $L(C)$. Then the local trivialization induces isomorphisms 
${\cal A}_{L(C), A_0} \sim G^{N_1}$ and  ${\cal G}_{L(C), A_0} \sim G^{N_0 - 1}$ where $N_1$ and $N_0$ are respectively the number of edges and vertices in $L(C)$. 
In this way a local trivialization puts $\frac{{\cal A}_{C\!\hbox{\footnotesize -flat}}}{{\cal G}_{M, \star}}$ in correspondence with $G^{N_1} / G^{N_0 - 1}$, where $G^{N_0 - 1}$ acts by the relevant adjoint action as in the gauge transformations of lattice gauge theory. 

%\begin{quote}
{\footnotesize 
We remark that in principle our parametrizations of 
${\cal A}_{C\!\hbox{\footnotesize -flat},A_0}$ and ${\cal G}_{C,A_0, \star}$ 
hold only inside the open set $U$ where $A_0$ is defined. 
We can choose an open cover of $M$ in which each open set is a union of cells in $C$, and proceed as above for each open set. Then we paste all the local parametrizations with the aid of some transition functions. This strategy leads to  definitions of partial gauge fixing of 
${\cal A}_{C\!\hbox{\footnotesize -flat}}$ and ${\cal G}_{C, \star}$ that hold in all of $M$. 
For any choice of transition functions we have a true partial gauge fixing; in the sense that the resulting space contains elements of all the gauge equivalence classes. Thus all the choices of transition functions are equivalent. This radical difference with the theory of smooth bundles is due to working with 
$\bar{\cal G}_{M, \star}$. Since this fact makes all the topological considerations almost trivial we will not be concerned about them in the rest of this article. 
}
%\end{quote}

Let us summarize the results stated above in the form of a lemma. 
\begin{lemma}
The following spaces are naturally isomorphic: 
\[
\frac{{\cal A}_{C\!\hbox{\footnotesize -flat}} }{ \bar{\cal G}_{M, \star}}
\sim 
\frac{{\cal A}_{C\!\hbox{\footnotesize -flat},A_0} }{ {\cal G}_{C,A_0, \star}}
\hbox{ and } 
\frac{{\cal A}_{L(C)} }{ {\cal G}_{L(C), b} }
\sim 
\frac{{\cal A}_{L(C), A_0} }{ {\cal G}_{L(C),A_0, b} } .
\] 
Additionally 
\[
{\rm Emb}_{L(C)}^{\star}:
\frac{{\cal A}_{C\!\hbox{\footnotesize -flat}} }{ \bar{\cal G}_{M, \star} }
\to 
\frac{{\cal A}_{L(C)} }{ {\cal G}_{L(C), b} }
\hbox{ and } 
{\rm Emb}_{L(C)}^{\star}:
{\cal A}_{C\!\hbox{\footnotesize -flat}, A_0} 
\to 
{\cal A}_{L(C), A_0} 
\]
are natural isomorphisms which are independent of the choice of representative embedding 
${\rm Emb}_{L(C)}: L(C) \to M$ (with ${\rm Emb}_{L(C)}(b) = \star$). 

Thus, 
\[
{\cal A}_{C\!\hbox{\footnotesize -flat}} / \bar{\cal G}_{M, \star} \sim 
G^{N_1(L(C))} / G^{N_0(L(C)) - 1} . 
\]
\end{lemma}
Our intuition about the lack of local degrees of freedom in $C$-flat connections 
is realized in the form of an identification between the space of $C$-flat connections modulo gauge transformations and a finite dimensional quotient space. 

%IN THE PART GAUGE FIX GAUGE AND DIFF INV ARE BROKEN. 
%THE LAST SECTION SOLVES THE PROBLEM

Within the partially gauge fixed construction 
both gauge invariance and diffeomorphism invariance are broken by the regularity assumptions. However, note that the original construction does not break gauge invariance, but still breaks diffeomorphism invariance. 
In a separate publication 
we will construct an ``extended" space of effective configurations which
admits a nontrivial action of the diffeomorphism group, and that after 
a partial gauge fixation 
yields our space ${\cal A}_{C\!\hbox{\footnotesize -flat}}$ \cite{K-flat}.

\subsection{Configuration spaces relevant for the continuum limit}
\label{SecContLim}
%CONTINUUM LIMIT IN OUR FRAMEWORK
A most important element in renormalization is the change to a coarser or finer  scale, and a limit in which the scale is the smallest possible (the continuum limit). 
Thus, in any extended notion of scale used to formulate effective theories 
there must be a relation that lets us know if 
a  measuring ``scale" is finer than another one, and the limit of the smallest ``scale" must make sense. 

Recall that the set of cellular decompositions of a manifold admits a partial order relation that tells us if a cellular decomposition is finer than another one ($\leq$), and that this partial order is directed in the direction of refinement towards ``the finest cellular decomposition" ($C\to M$)%
\footnote{
We remark that the directionality of the set of cellular decompositions holds in the piecewise linear and piecewise analytic categories, but not in the smooth category. In the extended framework (not partially gauge fixed with respect to the diffeomorphism symmetry) the directionality property holds also in the smooth category \cite{K-flat}. 
}. 
This partial order is preserved by our assignment of effective theories to cellular decompositions in the sense of ``being contained in" as 
subsets of 
$\bar{\cal A}_M$ or subsets of $\bar{\cal A}_M / \bar{\cal G}_{M,\star}$. 
(Later on we will use the notation 
 $i_C: {\cal A}_{C\!\hbox{\footnotesize -flat}} \to 
 \bar{\cal A}_M$ and 
 $i_C: {\cal A}_{C_1\!\hbox{\footnotesize -flat}} / \bar{\cal G}_{M,\star}
 \to \bar{\cal A}_M / \bar{\cal G}_{M,\star}$ for the inclusion maps.)
\begin{lemma}\label{nested}
Our assignment of effective gauge fields to cellular decompositions 
respects the partial order relation. 
Namely, $C_1 \leq C_2$ implies 
\begin{eqnarray*}
{\cal A}_{C_1\!\hbox{\footnotesize -flat}} \subseteq 
{\cal A}_{C_2\!\hbox{\footnotesize -flat}} 
\quad &,& \quad
{\cal A}_{C_1\!\hbox{\footnotesize -flat}, A_0} \subseteq 
{\cal A}_{C_2\!\hbox{\footnotesize -flat}, A_0} 
\\
{\cal A}_{C_1\!\hbox{\footnotesize -flat}}/\bar{\cal G}_{M,\star} &\subseteq &
{\cal A}_{C_2\!\hbox{\footnotesize -flat}}/\bar{\cal G}_{M,\star} 
\end{eqnarray*}
and $i_{C_1 \to C_2}$ 
will denote all the respective refining maps%
\footnote{
The injective map 
${\rm Emb}_{L(C_2)}^{\star}
 \circ i_{C_1 \to C_2}  \circ 
({\rm Emb}_{L(C_1)}^{\star}|_{{\cal A}_{C_1\!\hbox{\footnotesize -flat}, A_0}})^{-1}
: {\cal A}_{L(C_1), A_0} \to {\cal A}_{L(C_2), A_0}$, 
which is not really an inclusion, will also be denoted by $i_{C_1 \to C_2}$. 
Similarly, at the partially gauge fixed level we will use the same notation
for the maps induced by the refining. The ambiguity should be resolved by the context. 
}.  
\end{lemma}

Now let us introduce some configuration spaces that will be the kinematical basis of the continuum limit. 
First we present  the space of connections that are 
eventually $C$-flat (given the directed partial order of the cellular decompositions). 
This space can also be seen as the space of connections that are 
$C$-flat according to some cellular decomposition. 
\begin{definition}
\[
\tilde{\cal A}_M = \cup_C {\cal A}_{C\!\hbox{\footnotesize -flat}}. 
\]
\end{definition}
There are smaller configuration spaces labeled by a given triangulation $\Delta$ of $M$ that are also relevant for the continuum limit. 
These are constructed iterating the refining operation called barycentric subdivision (see the appendix), 
\begin{definition}
\[
\tilde{\cal A}_{M, \Delta} = \cup_n {\cal A}_{{\rm Sd}^n(\Delta)\!\hbox{\footnotesize -flat}}. 
\]
\end{definition}
We will see in the next section that $\tilde{\cal A}_{M, \Delta}$ can also be constructed as a projective limit, while $\tilde{\cal A}_M$ cannot. 

The interesting property is that while 
each ${\cal A}_{C\!\hbox{\footnotesize -flat}}$ captures very little of the information stored in $\bar{\cal A}_M$, both $\tilde{\cal A}_M$ and 
${\cal A}_{{\rm Sd}^n(\Delta)\!\hbox{\footnotesize -flat}}$ can be used to approximate any generalized connection arbitrarily well. 
\begin{theorem}
\[
\lim_{C \to M} {\cal A}_{C\!\hbox{\footnotesize -flat}} = 
\lim_{n \to \infty} {\cal A}_{{\rm Sd}^n(\Delta)\!\hbox{\footnotesize -flat}} = 
\bar{\cal A}_M
\]
in the sense that the subset of 
generalized connections composed by elements that are eventually in 
${\cal A}_{C\!\hbox{\footnotesize -flat}}$ 
(${\cal A}_{{\rm Sd}^n(\Delta)\!\hbox{\footnotesize -flat}}$) 
is 
$\tilde{\cal A}_M$ ($\tilde{\cal A}_{M, \Delta}$) 
which is dense in $\bar{\cal A}_M$. 
\end{theorem}
Proof. 

\noindent
While the restriction of two distinct cylindrical functions 
$f, g \in Cyl({\cal A}_M)$
to some ${\cal A}_{C\!\hbox{\footnotesize -flat}}$ may agree, 
it is a clear fact that if 
$f|_{\tilde{\cal A}_{M, \Delta}}= g|_{\tilde{\cal A}_{M, \Delta}}$ then $f= g$. 
In particular, if $f|_{\tilde{\cal A}_{M, \Delta}}= 0$ then $f$ is the zero of the algebra 
$Cyl({\cal A}_M)$. 

Now suppose that $\tilde{\cal A}_M$ is not dense in  $\bar{\cal A}_M$. 
Then there is $A \in\bar{\cal A}_M$ and a whole neighborhood of it ${\cal N}_A$ 
such that ${\cal N}_A \subset \bar{\cal A}_M - \tilde{\cal A}_M$. 
Thus, there is a non zero continuous function $f \in Cyl({\cal A}_M)$ such that 
$f|_{\bar{\cal A}_M - {\cal N}_A}= 0$. In particular our assumption implies that there is a $f \in Cyl({\cal A}_M)$ which 
$f|_{\tilde{\cal A}_M}= 0$  while it is not the zero of $Cyl({\cal A}_M)$. 
The contradiction  implies that 
$\tilde{\cal A}_M$ is indeed dense in  $\bar{\cal A}_M$. 
\\$\Box$

Although $\tilde{\cal A}_M$ and its subset $\tilde{\cal A}_{M, \Delta}$ are both dense 
inside $\bar{\cal A}_M$, 
in the measure theoretical sense they are both small subsets of $\bar{\cal A}_M$. 
The formal statement at the gauge invariant level is the following. 
\begin{theorem}
As subsets of the space $\bar{\cal A}_M / \bar{\cal G}_{M,\star}$ equipped with the measure $\mu_{AL}$,  
$\tilde{\cal A}_{M, \Delta}/ \bar{\cal G}_{M,\star}$ and 
$\tilde{\cal A}_M / \bar{\cal G}_{M,\star}$
are thin sets. 
\end{theorem}
Proof. 

\noindent
The proof is a simple adaptation of Thiemann's 
theorem on the support of 
the Ashtekar-Lewandowski measure \cite{thiemann} (Theorem I.212). 

We will construct a map 
\[
h^s: \bar{\cal A}_M / \bar{\cal G}_{M,\star} \to G^{[0,1]}
\]
that will let us study ``a piece of $\bar{\cal A}_M$." To make this possible we consider $G^{[0,1]}$ equipped with a different topology and measure theoretical structure than the usual ones. For the moment consider $G^{[0,1]}$ as a set; its structure will be defined below to be compatible with the structure of $\bar{\cal A}_M$ trough the map $h^s$. 

Consider a one parameter family of loops based at $\star \in M$, 
$s(t) \in {\cal P}_{\star}$, $t \in [0, 1]$, 
such that $s(0) = {\rm id}_{\star}$ and $s(1)$ is a holonomicaly non trivial loop. Generalized connections assign group elements to loops, 
$A(s(t)) \in G$. The latter can be seen as a function from $[0, 1]$ to $G$, or equivalently an element of $G^{[0,1]}$. Thus, a one parameter family of loops $s(t)$ induces a map from 
$ \bar{\cal A}_M$ to $G^{[0,1]}$. This is our map $h^s: \bar{\cal A}_M \to G^{[0,1]}$. 

Given a choice of finitely many points $t_i \in [0, 1]$ 
consider it as an assignment from $G^{[0,1]}$ to some $G^n$. Thus 
continuous functions from $f: G^n \to \C$ induce {\em cylindrical functions} 
$f_{\{ t_i \}}: G^{[0,1]} \to \C$. Clearly the pull back of these functions by $h^s$ are cylindrical functions of $\bar{\cal A}_M$. We endow $G^{[0,1]}$ with the weakest topology  that makes cylindrical functions continuous and also consider these cylinder functions the basis of its measure theoretical structure. Additionally, the push forward of the Ashtekar-Lewandowski measure is the natural  homogeneous measure on $G^{[0,1]}$ endowed with this structure. 

Consider the subset $D \subset G^{[0,1]}$ defined by those functions 
$h: [0,1] \to G$ that are nowhere continuous {\em according to the usual topologies} of $[0,1]$ and $G$. 

Thiemann proves that $G^{[0,1]} - D$ is contained in a measure zero set of $G^{[0,1]}$ with respect to the homogeneous measure \cite{thiemann}. 

Our result follows from the fact that 
$h^s(\tilde{\cal A}_{M, \Delta}/ \bar{\cal G}_{M,\star} ) \subset G^{[0,1]} - D$ and 
$h^s(\tilde{\cal A}_{M}/ \bar{\cal G}_{M,\star} ) \subset G^{[0,1]} - D$. 
\\$\Box$ 

%%%%%%

Given that our spaces of effective configurations are nested in the sense of lemma \ref{nested}, each ${\cal A}_{C\!\hbox{\footnotesize -flat}}$ can be used to regularize (approximate) any 
cylindrical function from the continuum or from a finer cellular decomposition. 
This regularization is naturally performed by the pull back of the inclusion maps 
$i_C: 
{\cal A}_{C_1\!\hbox{\footnotesize -flat}} / \bar{\cal G}_{M,\star}
 \to 
 \bar{\cal A}_M / \bar{\cal G}_{M,\star}$ 
 or 
$i_{C_1 \to C_2}: 
{\cal A}_{C_1\!\hbox{\footnotesize -flat}}/\bar{\cal G}_{M,\star} 
\to 
{\cal A}_{C_2\!\hbox{\footnotesize -flat}}/\bar{\cal G}_{M,\star}$. 

Now we will define measures in $\bar{\cal A}_M$  as 
the continuum limit of effective measures. 
\begin{definition}[Continuum limit of effective theories]\label{contlim}
Consider a collection of 
measures $\{ \mu_C \}$ (or $\{ \mu_n \}$): 
one measure in each of the spaces of effective configurations 
$\{ {\cal A}_{C\!\hbox{\footnotesize -flat}} \}$ 
(or $\{ {\cal A}_{{\rm Sd}^n(\Delta)\!\hbox{\footnotesize -flat}} \}$). 

When the measures converge, the continuum limit measure $\mu_M$, will be a measure in 
$\bar{\cal A}_M$ defined by its action on cylindrical functions as follows: 
\[
\mu_M (f) \doteq \lim_{C \to M} \mu_C (i_C^{\star} f)
\] 
or 
\[
\mu_M (f) \doteq \lim_{n \to \infty} \mu_n (i_C^{\star} f)
\] 
for any cylindrical function $f\in Cyl({\cal A}_M)$. 
\end{definition}
Clearly the continuum limit of effective measures as $C \to M$ is a much stronger condition than the limit using a sequence of refinements of a given triangulation. 
The limit $C \to M$ needs that an effective measure be defined for any $C$ and that different refinement sequences of effective measures have the same limit. 
If our objective is to define measures in the continuum, we should use the limit $n \to \infty$ because it is more economical, but the limit $C \to M$ has the advantage of being manifestly independent of any choice of ``discretization" (refining sequence). 

The measures in finer and coarser effective theories should be related if they describe the same physical system. In section \ref{coarsegrainmeasures} we will show how a finer measure is coarse grained to produce a measure for a coarser effective theory. 
If in a given refining sequence the measures 
at coarser scales are constructed by coarse graining finer measures, we say that the measures are projectively compatible. In all these cases the limit 
$n \to \infty$ 
of our previous definition exists, 
but it is trivial in the sense that starting from a high enough $n_0$ for any $n\geq n_0$ $\mu_n (i_C^{\star} f)$ is independent of $n$. 
The known examples of measures that can be constructed 
in this way include the Ashtekar-Lewandowski measure (the homogeneous measure) and distributional measures peaked on flat connections. 

Our objective is to 
emulate the Lattice Gauge Theory procedure to find a theory in the continuum, which is much more powerful. 
There are many examples of 
collections $\{ \mu_n \}$ which are not projectively compatible, but as $n$ increases they become more and more compatible in the sense of the definition given above. It is easy to exhibit examples of nontrivial convergence for sigma models, and one can also construct them for gauge fields.

When the continuum limit exist, the measure $\mu_M$ can be coarse grained to define measures in all the spaces ${\cal A}_{{\rm Sd}^n(\Delta)\!\hbox{\footnotesize -flat}}$. Clearly this sequence of measures will be projectively compatible. These measures should be considered as ``completely renormalized." In general they differ from the original effective measure used to construct the continuum limit measure, 
but in the sense of the limit defined above they 
differ less and less as we approach the continuum limit. 

In a restricted version of our framework 
where we consider only regular cellular decompositions 
we can simply declare that the effective measures are the ones used in lattice gauge theory (by a renormalization group procedure). 
From this point of view there is a whole body of evidence supporting the existence of nontrivial physically interesting measures. A more detailed study of the relation between our framework and standard lattice gauge theory is needed. 

The question arises of which measures on $\bar{\cal A}_M$ can be defined as a continuum limit. This question and many others are outside of the scope of this article. 
A detailed measure theoretical study of this continuum limit is certainly needed. 

\section{Coarse graining and \\
fiber bundle structure}

Consider a situation in which we have two effective theories, one of which is based on finer knowledge of the system. The coarser and the finer effective theories should be related by a renormalization procedure which adjusts the coupling constants 
to account for the bulk effect of averaging out the degrees of freedom of the finer theory which are considered as fluctuations over the degrees of freedom relevant for the coarser theory. 

In the previous section we defined spaces of effective connections. The space of connections related to a finer theory contains that of a coarser theory. 
Physically this will let us treat the degrees of freedom used to describe coarser configurations as background degrees of freedom. 

In this section we define a coarse graining procedure based on a choice of fibration which organizes the remaining  degrees of freedom of the finer theory as fluctuations over each of the coarser (background) configurations. This fibration and Fubini's theorem tell us how to integrate out the fluctuations; a measure for the coarser theory is induced by coarse graining the measure for the finer theory. 

%Definition of the coarse graining map \pi 
\begin{definition}[Coarse graining maps]
A natural coarse graining in our setting is induced by a choice of representative embedding 
${\rm Emb}_{L(C)}: L(C) \to M$. 
\[
\pi_C \doteq ({\rm Emb}_{L(C)}^{\star}|_{{\cal A}_{C\!\hbox{\footnotesize -flat}}/\bar{\cal G}_{M,\star}})^{-1} \circ {\rm Emb}_{L(C)}^{\star} 
: \bar{\cal A}_M/\bar{\cal G}_{M,\star} \to {\cal A}_{C\!\hbox{\footnotesize -flat}}/ \bar{\cal G}_{M,\star} , 
\]
where ${\rm Emb}_{L(C)}^{\star} : {\cal A}_{C\!\hbox{\footnotesize -flat}}/\bar{\cal G}_{M,\star}  \to 
{\cal A}_{L(C)} / {\cal G}_{L(C), b}$ is an isomorphism. 
\[
\pi_C \doteq ({\rm Emb}_{L(C)}^{\star} |_{{\cal A}_{C\!\hbox{\footnotesize -flat}, A_0}})^{-1} \circ P_{A_0} \circ {\rm Emb}_{L(C)}^{\star} 
: \bar{\cal A}_M \to {\cal A}_{C\!\hbox{\footnotesize -flat}, A_0} , 
\]
where 
$P_{A_0}: {\cal A}_{L(C)}  \to 
{\cal A}_{L(C), A_0}$ is the partial gauge fixation and 
${\rm Emb}_{L(C)}^{\star} : {\cal A}_{C\!\hbox{\footnotesize -flat}, A_0}  \to {\cal A}_{L(C), A_0}$ is an isomorphism. 
Given $C_1 \leq C_2$ 
\[
\pi_{C_2 \to C_1} \doteq 
\pi_{C_1} |_{{\cal A}_{C_2\!\hbox{\footnotesize -flat}} / \bar{\cal G}_{M,\star}} 
: {\cal A}_{C_2\!\hbox{\footnotesize -flat}}/ \bar{\cal G}_{M,\star} \to 
{\cal A}_{C_1\!\hbox{\footnotesize -flat}} / \bar{\cal G}_{M,\star}.
\]
\[
\pi_{C_2 \to C_1} \doteq \pi_{C_1} |_{{\cal A}_{C_2\!\hbox{\footnotesize -flat}, A_0}} 
: {\cal A}_{C_2\!\hbox{\footnotesize -flat}, A_0} \to 
{\cal A}_{C_1\!\hbox{\footnotesize -flat}, A_0} .
%\]
%These coarse graining maps have the same name because they are compatible %and they are projections
\footnote{
When working on the spaces ${\cal A}_{L(C_i)}$ the relevant coarse graining map is $\pi_{C_2 \to C_1} = {\rm Emb}_{1, 2}^{\star}$, where ${\rm Emb}_{1, 2}: L(C_1) \to L(C_2)$ represents a $C_2$-equivalence class of representative embeddings ${\rm Emb}_{L(C_1)}$. 
Note that this map is not a projection. 
The ambiguity in our notation should be resolved by the context. 
} 
\]
\end{definition}

In the rest of this section we will treat three different subjects in corresponding subsections. In the first subsection we will give a minimal example of our spaces of effective configurations and coarse graining maps; in addition we use the example to 
motivate the general result of the following subsection. In the second subsection we present a theorem describing the fiber bundle structure induced by our coarse graining maps. In the final subsection we then study the effect of coarse graining on measures; in particular we treat the issue of renormalization prescriptions: conditions that relate the measures on different effective theories ``asking them to describe the same physical system."

\subsection{A minimal example}
%Simple examples (Yorch) and regular lattice mentioning generic and 
%non generic embeddings
%Mention that usually regularization assumes generic coarse graining. 
Here we present a minimal example and 
use it to begin our exploration of the general properties of the coarse graining map. 
%The example of a regular two dimensional lattice is presented in appendix 
%\ref{B}. 
For  concreteness we present it at the partially gauge fixed level. 

Consider a closed disc (a two dimensional closed ball) with the cellular decompositions $C_1,C_2$ described in the figure below. These cellular decompositions induce the auxiliary lattices $L(C_1),L(C_2)$ also depicted in the figure. 
\begin{figure}[!ht]
\centerline {
\includegraphics[width=6cm]{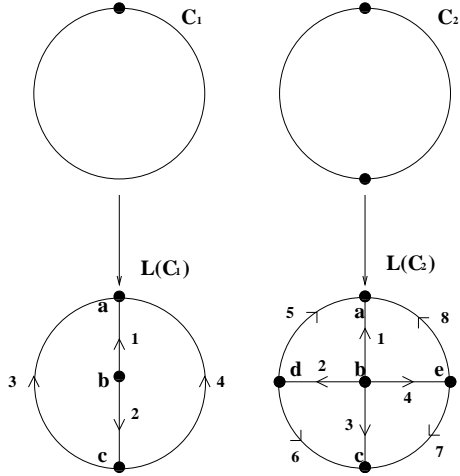}\hspace{2.5cm}
}
\caption{Two cellular decompositions of the disc, $C_1 \leq C_2$.}\label{F;3}
\end{figure}
In the parametrizations described in the previous section we can write \\
$A\in  {\cal A}_{L(C_1), A_0}$ as
$A=(g_1,g_2,g_3,g_4)\in G^4$, \\
$g\in {\cal G}_{L(C_1), A_0}$ as 
$g=(g_a,g_b,g_c)\in G^3$, \\
$A'\in {\cal A}_{L(C_2), A_0}$ as 
$A'=(g'_1,g'_2,g'_3,g'_4,g'_5,g'_6, g'_7,g'_8)\in G^8$ and \\
$g'\in {\cal G}_{L(C_2), A_0}$ as 
$g'=(g'_a,g'_b,g'_c, g'_d, g'_e)\in G^5$. \\ 
We will also use the subgroup of gauge transformations that are the identity at a vertex, our notation will be \\
$g\in {\cal G}_{L(C_1), A_0, b}$ as 
$g=(g_a,g_c)\in G^2$ and \\
$g'\in {\cal G}_{L(C_2), A_0, b}$ as 
$g'=(g'_a,g'_c, g'_d, g'_e)\in G^4$. \\
The corresponding connections modulo gauge can be written as \\
$[A]\in  {\cal A}_{L(C_1), A_0}/{\cal G}_{L(C_1), A_0,b}$ as
$A=(g_{l1},g_{l2})\in G^2$ and \\
$[A']\in  {\cal A}_{L(C_2), A_0}/{\cal G}_{L(C_2), A_0,b}$ as
$[A']=(g'_{l'1},g'_{l'2}, g'_{l'3},g'_{l'4})\in G^4$, \\
where $l1=    2^{-1} \circ 3^{-1} \circ 1$ and 
$l2= 1^{-1} \circ 4 \circ 2$ are loops in $L(C_1)$ based at $b$, and 
$l'1=    2^{-1} \circ 5^{-1} \circ 1$, 
$l'2=    3^{-1} \circ 6 \circ 2$, 
$l'3=    4^{-1} \circ 7^{-1} \circ 3$ and 
$l'4= 1^{-1} \circ 8 \circ 4$ are loops in $L(C_2)$ based at $b$.

The maps induced by refining (the inclusion of $C$-flat connections) on connections $i_{C_1 \to C_2}: {\cal A}_{L(C_1), A_0} \to {\cal A}_{L(C_2), A_0}$ 
and gauge transformations 
$i_{C_1 \to C_2}: {\cal G}_{L(C_1), A_0} \to {\cal G}_{L(C_2), A_0}$ 
are 
\begin{eqnarray*}
i_{C_1 \to C_2}(A)&=&(g_1,g_2,g_2,g_2,g_3,{\rm id},{\rm id},g_4),\\
i_{C_1 \to C_2}(g)&=&(g_a,g_b,g_c,g_c,g_c). 
\end{eqnarray*}
In addition, the three different coarse graining maps 
depending on the choice of
embedding ${\rm Emb}_{1, 2}: L(C_1) \to L(C_2)$ are
\begin{eqnarray*}
\pi_1(A')&=&(g'_1,g'_2,g'_5,g'_8(g'_7)^{-1}g'_6),\\
\pi_1(g')&=&(g'_a,g'_b,g'_d),\\
\pi_2(A')&=&(g'_1,g'_3,g'_5(g'_6)^{-1},g'_8(g'_7)^{-1}),\\
\pi_2(g')&=&(g'_a,g'_b,g'_c),\\
\pi_3(A')&=&(g'_1,g'_4,g'_5(g'_6)^{-1}g'_7,g'_8) \\
\pi_3(g')&=&(g'_a,g'_b,g'_e). 
\end{eqnarray*}

It is easy to verify that the maps $i_{C_1 \to C_2}$ and $\pi_i$ descend to the quotient by gauge transformations since 
$i_{C_1 \to C_2}(g(A)) = \\
i_{C_1 \to C_2}(g) (i_{C_1 \to C_2}(A))$, and 
$\pi_i(g'(A')) = \pi_i(g') (\pi_i(A'))$. 

We will work out the example using the projection $\pi_1$. 
The other projections $\pi_i$ yield the same structures. Our study will 
show that in this example 
$({\cal A}_{L(C_2), A_0},\pi_1, {\cal A}_{L(C_1), A_0})$ 
and \\
$({\cal A}_{L(C_2), A_0}/ {\cal G}_{L(C_2), A_0,b},\pi_1, 
{\cal A}_{L(C_1), A_0}/{\cal G}_{L(C_2), A_0,b})$ 
are principal fiber bundles with structure group $G^4$ and $G^2$ respectively. In addition, they have a preferred global section induced by the 
refining maps 
(inclusion of $C$-flat connections). 

On the other hand, 
$({\cal A}_{L(C_2), A_0}/ {\cal G}_{L(C_2), A_0},\pi_1, 
{\cal A}_{L(C_1), A_0}/{\cal G}_{L(C_2), A_0})$ 
has the structure of a bundle where neither the total space nor the base are manifolds and where most fibers are homeomorphic to $G^2$, the only exceptions being the fibers over non generic connections modulo gauge, which are not typical. For example, the fiber over the flat connection is 
the quotient space generated by two copies of the gauge group modulo the adjoint action, $G^2 / {\rm Ad} G$. We will not study this space 
directly; the most illuminating fact about it is that it is the quotient of the principal fiber bundle \\
$({\cal A}_{L(C_2), A_0}/ {\cal G}_{L(C_2), A_0,b},\pi_1, 
{\cal A}_{L(C_1), A_0}/{\cal A}_{L(C_2), A_0,b})$ 
by $G$, that is a finite dimensional Lie group.

Now we start with the study of the typical fiber of 
\[
({\cal A}_{L(C_2), A_0},\pi_1, {\cal A}_{L(C_1), A_0}).
\] 
Given any $A_0=(g^\circ_1,g^\circ_2,g^\circ_3,g^\circ_4)\in
{\cal A}_{L(C_1), A_0}$ the fiber over it can be parametrized as 
\[
\pi^{-1}_1(A_0)=\{ (g^\circ_1,
g^\circ_2,g_3,g_4,g^\circ_3,g_7(g_8)^{-1}g^\circ_4,g_7,g_8)\}.
\]

Then all the fibers of $({\cal A}_{L(C_2), A_0},\pi_1, {\cal A}_{L(C_1), A_0})$ are homeomorphic to $G^4$.

A bit of graphical thinking will let us understand this example and generalize the result. 
At the non gauge invariant level $G^4$ appears because 
we have eight unknowns (the number of edges in $L(C_2)$) and four restrictions 
(the number of edges in $L(C_1)$). 
Once we have found one solution to the conditions, we can generate more solutions in two ways: 
(i) modifying only holonomies along edges in the graph 
$L(C_2) - {\rm Emb}_{1, 2} (L(C_1))$, 
(ii) modifying holonomies along edges of $L(C_2)$ covered by 
${\rm Emb}_{1, 2} (L(C_1))$ without modifying the induced holonomies in the edges of $L(C_1)$. 

In our example 
(i) corresponds to modifying the components of $A'$ assigned to edges $3, 4$ of $L(C_2)$. 
Whereas (ii) must be achieved by 
modifications that involve the components of $A'$ assigned to edges $6, 7, 8$ of $L(C_2)$ without modifying the holonomy along the image by ${\rm Emb}_{1, 2} $ of edge $4$ of $L(C_1)$. 
Of course, these modifications can be of the form of ``gauge transformations" acting on vertices $c,e$ of $L(C_2)$. Moreover, these transformations of the 
``gauge transformation" style together with  type (i) transformations 
generate all the possible modifications of solutions to the restrictions. 

Then, consider the following 
$G^2\times G^2$ left action on ${\cal A}_{L(C_2), A_0}$ 
\begin{eqnarray*}
&\mathcal{F}_{\nu_3, \nu_4; \nu_c, \nu_e} (g_1^\circ,\ldots,g_8^\circ)=&\\
&(g^\circ_1,g^\circ_2,\nu_cg^\circ_3\nu_3^{-1},
\nu_eg^\circ_4\nu_4^{-1},g^\circ_5,\nu_cg^\circ_6,\nu_cg^\circ_7\nu_e^{-1},g^\circ_8\nu_e^{-1}).&
\end{eqnarray*}
In our notation type (i) transformations are parametrized by $(\nu_3, \nu_4)\in G^2$ and type (ii) transformations are parametrized by  $(\nu_c, \nu_e)\in G^2$. 

Our argument in the previous paragraph implies that 
the map $\mathcal{F}_\nu$ preserves fibers, 
$\pi_1(A_0')=\pi_1(\mathcal{F}_\nu (A_0'))$. 
Furthermore, it gives a bijection between each fiber and $G^4$. Then
$({\cal A}_{L(C_2), A_0},\pi_1,{\cal A}_{L(C_2), A_0})$ is a $G^4$ principal
fiber bundle.

Now we would like to describe the typical fiber of 
\[
({\cal A}_{L(C_2), A_0}/ {\cal G}_{L(C_2), A_0,b},\pi_1, 
{\cal A}_{L(C_1), A_0}/{\cal G}_{L(C_1), A_0,b}) ; 
\] 
the procedure will follow our previous analysis closely. 
Given any $[A_0]=(g^\circ_{l1},g^\circ_{l2})\in {\cal A}_{L(C_1), A_0}/{\cal G}_{L(C_1), A_0,b}$ the fiber over it can be parametrized as 
\[
\pi^{-1}_1([A_0])=\{ (g^\circ_{l1},
g'_{l'2}, g'_{l'3}, 
g^\circ_2 g'^{-1}_{l'3} g'^{-1}_{l'2})\}.
\]
Then all the fibers of $({\cal A}_{L(C_2), A_0}/ {\cal G}_{L(C_2), A_0,b},\pi_1, 
{\cal A}_{L(C_1), A_0}/{\cal G}_{L(C_1), A_0,b})$ are homeomorphic to $G^2$.

Consider the following 
$G^2$ left action on ${\cal A}_{L(C_2), A_0}/ {\cal G}_{L(C_2), A_0,b}$ 
\begin{eqnarray}\label{g-transf-type}
&\tilde{\mathcal{F}}_{\nu_1, \nu_2} (g'_{l'1},g'_{l'2}, g'_{l'3},g'_{l'4})=&\\
&(g'_{l'1},
\nu_1 g'_{l'2}, 
\nu_2 g'_{l'3} \nu_1^{-1},
g'_{l'4} \nu_2^{-1}).&
\end{eqnarray}
Note that it is the action induced by $\mathcal{F}_\nu$ at the gauge invariant level. 

Clearly the map $\tilde{\mathcal{F}}_\nu$ 
preserves fibers, 
$\pi_1([A_0'])=\pi_1(\tilde{\mathcal{F}}_\nu ([A_0']))$. 
Furthermore, it gives a bijection between each fiber and $G^2$. Then \\
$({\cal A}_{L(C_2), A_0}/ {\cal G}_{L(C_2), A_0,b},\pi_1, 
{\cal A}_{L(C_1), A_0}/{\cal G}_{L(C_1), A_0,b})$ is a $G^2$ principal
fiber bundle.

\subsection{Fiber bundle structure}
%Theorem about fiber bundle structure
\begin{theorem}
Given a cellular decomposition and a refinement of it, $C_1 \leq C_2$, 
the triples 
$({\cal A}_{C_2\!\hbox{\footnotesize -flat}}/{\cal G}_{C_2, \star},
\pi_{C_2 \to C_1}, 
{\cal A}_{C_1\!\hbox{\footnotesize -flat}}/{\cal G}_{C_1, \star})$ 
and \\
$({\cal A}_{C_2\!\hbox{\footnotesize -flat}, A_0},
\pi_{C_2 \to C_1}, 
{\cal A}_{C_1\!\hbox{\footnotesize -flat}, A_0})$ 
are principal fiber bundles with a preferred global section induced by the 
refining maps 
(inclusion of $C$-flat connections). 
\end{theorem}
Proof. 

\noindent
The proof of the general case is entirely analogous to our treatment of the example. 
Here we present it only at the gauge invariant level. 

${\cal A}_{C_2\!\hbox{\footnotesize -flat}}/{\cal G}_{C_2, \star}$  
-- or equivalently, ${\cal A}_{L(C_2)} / {\cal G}_{L(C_2), b}$ --
is parametrized by 
$G^{N_1(L(C_2)) - (N_0(L(C_2)) - 1)}$. 

Once the base space point is fixed in the parametrization of \\
${\cal A}_{C_1\!\hbox{\footnotesize -flat}}/{\cal G}_{C_1, \star}$ as 
$G^{N_1(L(C_1)) - (N_0(L(C_1)) - 1)}$, the locus of the fiber over it is found by solving $N_1(L(C_1)) - (N_0(L(C_1)) - 1)$ conditions on the variables 
$G^{N_1(L(C_2)) - (N_0(L(C_2)) - 1)}$. 

When a solution has been found for 
these equations, one can find all the other solutions trough a \\
$G^{[N_1(L(C_2)) - (N_0(L(C_2)) - 1)]- [N_1(L(C_1)) - (N_0(L(C_1)) - 1)]}$ action, 
$\tilde{\mathcal{F}}_\nu$, 
of the \\
gauge transformation type described in \ref{g-transf-type}. 

Clearly the map $\tilde{\mathcal{F}}_\nu$ 
preserves fibers and gives a bijection between each fiber and 
$G^{[N_1(L(C_2)) - (N_0(L(C_2)) - 1)]- [N_1(L(C_1)) - (N_0(L(C_1)) - 1)]}$. 
\\$\Box$

%Corollary on embeddings of effective H-F algebras. 
Since the $C$-effective theory is isomorphic to a lattice gauge theory, it is clear that configuration observables correspond to holonomies and momentum observables correspond to left invariant vector fields. The set of these observables is an algebra under a Poisson bracket product and is called the holonomy flux algebra, $\hbox{H-F}(C)$. 

The pull back of the coarse graining map takes holonomies from $\hbox{H-F}(C_1)$ to $\hbox{H-F}(C_2)$ and the 
left invariant push forward of the refining map can be used to bring fluxes from 
$\hbox{H-F}(C_1)$ to $\hbox{H-F}(C_2)$. We call this map $\widehat{\pi_{C_2 \to C_1}^{\star}}$. 
It turns out that 

\begin{corollary}
If $C_1 \leq C_2$ 
\[
\widehat{\pi_{C_2 \to C_1}^{\star}} : \hbox{H-F}(C_1) \to \hbox{H-F}(C_2)
\]
is a $\star$-algebra embedding. 
\end{corollary}
The proof of this statement, a detailed study of coarse graining within the algebraic approach and 
phase space effective theories will be treated elsewhere.

\subsection{Coarse graining effective measures}\label{coarsegrainmeasures}

In subsection \ref{SecContLim} we defined a construction of measures in $\bar{\cal A}_M$ as a continuum limit of effective measures on the spaces 
${\cal A}_{C\!\hbox{\footnotesize -flat}}$. 
If $C_1 \leq C_2$, the effective measures $\mu_{C_1}$ and $\mu_{C_2}$ 
have to be related by coarse graining (at least approximately) 
because they define effective theories 
for the same physical system. 
Assume that we have chosen a projection map $\pi_{C_2 \to C_1}$, which amounts to having chosen certain degrees of freedom on ${\cal A}_{C_2\!\hbox{\footnotesize -flat}}$ as fluctuations over the background configurations 
$i_{C_1 \to C_2} {\cal A}_{C_1\!\hbox{\footnotesize -flat}}$. 
\begin{definition}[Exact renormalization prescription] 
\[
(\pi_{C_2 \to C_1})_{\star} \mu_{C_2} = \mu_{C_1} .
\]
\end{definition}
The definition means that for any cylindrical function $f$ of ${\cal A}_{C_1\!\hbox{\footnotesize -flat}}$ we have 
$\int_{{\cal A}_{C_1\!\hbox{\footnotesize -flat}}} f d\mu_{C_1} = 
\int_{{\cal A}_{C_2\!\hbox{\footnotesize -flat}}} 
\pi_{C_2 \to C_1}^{\star} f d\mu_{C_2} $. 
Its physical interpretation is that any scale $C_1$ observable 
can either be measured by $\mu_{C_1}$ on the space $ {\cal A}_{C_1\!\hbox{\footnotesize -flat}}$ or be ``observed at scale $C_2$" as a rather coarse function and measured by 
$\mu_{C_2}$ on the space $ {\cal A}_{C_2\!\hbox{\footnotesize -flat}}$ 
{\em producing exactly the same results}.

Given any refining sequence $C1 \leq C_2 \leq ... \leq C_n  \leq ... $ there are choices of projections that make the corresponding exact renormalization prescriptions {\em compatible}. This is because the following lemma holds. 
\begin{lemma}
Given any three cellular decompositions related by refinement 
$C_1 \leq C_2 \leq C_3$ there are choices of embeddings 
${\rm Emb}_{i, j}: L(C_i) \to L(C_j)$ which induce projections 
$\pi_{C_j \to C_i}= {\rm Emb}_{i, j}^{\star}: 
{\cal A}_{L(C_j)}  \to {\cal A}_{L(C_i)}$
that make the following triangle diagram commute. 

%Triangle Diagram
{
\unitlength=1mm 
\special{em:linewidth 0.4pt}
\linethickness{0.4pt} 

\begin{center}
\begin{picture}(106.00,30.00) \thicklines
\put(38.00,25.00){\vector(1,0){32.00}}
\put(32.00,20.00){\vector(1,-1){15.00}}
\put(60.00,5.00){\vector(1,1){15.00}}
\put(53.00,29.00){\makebox(0,0)[cc]{
$\pi_{C_3 \to C_1}$
}}
\put(75.00,11.00){\makebox(0,0)[cc]{{
$\pi_{C_3 \to C_2}$
}}}
\put(32.00,11.00){\makebox(0,0)[cc]{
$\pi_{C_2 \to C_1}$
}}
\put(30.00,25.00){\makebox(0,0)[cc]
{{
${\cal A}_{L(C_1)}$
}}}
\put(78.00,25.00){\makebox(0,0)[cc]
{{
${\cal A}_{L(C_3)}$
}}}
\put(53.00,2.00){\makebox(0,0)[cc]
{{
${\cal A}_{L(C_2)}$
}}}
\end{picture}
\end{center}
}

\end{lemma}
Proof. 

\noindent
Fix ${\rm Emb}_{1, 2}: L(C_1) \to L(C_2)$  and 
${\rm Emb}_{2, 3}: L(C_2) \to L(C_3)$. Clearly 
${\rm Emb}_{2, 3} \circ {\rm Emb}_{1, 2}: L(C_1) \to L(C_3)$ 
sends holonomicaly independent paths into 
holonomicaly independent paths, and it also represents a 
$C_3$-equivalence class of embeddings 
${\rm Emb}_{L(C_1)}: L(C_1) \to M$ as required. 
\\$\Box$ 

In particular one can choose compatible exact renormalization prescriptions for the refining sequence 
$\Delta \leq {\rm Sd}(\Delta) \leq ... \leq {\rm Sd}^n(\Delta) \leq ...$.  In this case 
a solution of such chain of conditions would yield 
a sequence of projectively compatible measures $\{ \mu_n \}$. Thus, the continuum limit \ref{contlim} would exist and define a measure $\mu_M$ on 
$\bar{\cal A}_M$. 

It is important to remark that a 
collection of compatible embeddings that makes the exact renormalization prescriptions compatible can also be used to define the projective limit more commonly used in loop quantization and that the results are compatible in the following sense. 
\begin{theorem}
Given a family of compatible projections \\
$\{ \pi_{n+1 \to n}: 
{\cal A}_{{\rm Sd}^{n+1}(\Delta)\!\hbox{\footnotesize -flat}} \to 
{\cal A}_{{\rm Sd}^n(\Delta)\!\hbox{\footnotesize -flat}} \}$
the projective limit of the spaces ${\cal A}_{{\rm Sd}^n(\Delta)\!\hbox{\footnotesize -flat}}$ is $\tilde{\cal A}_{M, \Delta}$. 

In addition, the collection of projectively compatible measures $\{ \mu_n \}$ defines a measure $\tilde{\mu}_M$ on $\tilde{\cal A}_{M, \Delta}$ which is 
compatible with our continuum limit measure in the sense that 
\[
i_{\star} \tilde{\mu}_M = \mu_M ,
\]
where $i: \tilde{\cal A}_{M, \Delta} \to \bar{\cal A}_M$ is the inclusion map. 
\end{theorem}
The proof of this theorem is a simple corollary of definitions. 
Of course 
the continuum limit in these cases 
is trivial in the sense that starting from a high enough $n_0$ for any $n\geq n_0$ $\mu_n (i_C^{\star} f)$ is independent of $n$. 

Our continuum limit of measures significantly extends the projective limit and this extension is of interest for physical applications. We say that because there are 
many examples of 
collections $\{ \mu_n \}$ which are not projectively compatible, but as $n$ increases they become more and more compatible in 
a way that makes the continuum limit exist. 
More significantly, 
the construction of measures by standard lattice gauge theory can be imported 
to a restricted version of our framework 
where we consider only regular cellular decompositions. 
From this point of view there is a whole body of evidence supporting the existence of nontrivial physically interesting measures. 

In standard lattice gauge theory the used renormalization prescriptions are much weaker that our exact renormalization prescription: they ask only that some correlation functions 
chosen by their physical importance 
be preserved by coarse graining. Also, the allowed measures 
in LGT are of a very constrained type; only a few coupling constants completely specify their measures. 
With a smaller space of allowed measures and weaker renormalization prescriptions they are able to generate a flow in the space of coupling constants, which flows in the direction of refinement. 
We have followed and implemented their ideas in a few examples of our framework \cite{Ising}.

What about compatible renormalization prescriptions relating the effective theories assigned to all the cellular decompositions? 
Consider two refining sequences 
that share a cellular decomposition $C$. 
A choice of compatible projections for a refining sequence completely determines the embedding map ${\rm Emb}_{L(C)}: L(C) \to M$. Thus, 
if one chooses independently the two families of compatible projections, it could happen that the induced embedding ${\rm Emb}_{L(C)}: L(C) \to M$ 
may be different for the two sequences. This would mean that the projection from $\bar{\cal A}_M$ to 
${\cal A}_{C\!\hbox{\footnotesize -flat}}$ would depend on the coarse graining route. 
If the coarse graining between two effective theories depends on the coarse graining route, there would be one exact renormalization prescription 
per path and they would not be compatible with each other. 
Weather 
there are choices of embeddings that avoid 
``coarse graining route dependence" or there aren't, and many other issues about 
renormalization prescriptions will be treated in \cite{Ising}. 

Again we stress that the lattice gauge theory experience tells us that non exact renormalization prescriptions are more interesting than exact ones. Then, the more relevant question would be about compatibility of (non exact) renormalization prescriptions. In this context it is known that coarse graining does depend on the coarse graining route. A familiar example is performing Migdal-Kadanoff blocking in two dimensions first in the $x$ direction and then in the $y$ direction, or reversing the order of the blocking. In this case the issue is resolved by symmetry considerations; a treatment of symmetries in our framework is presented 
in \cite{K-flat}.

\section{Action of ``diffeomorphisms"}
\label{gauge+diff}

Clearly general fiber bundle maps do not leave 
the spaces of $C$-flat connections invariant, but their action is very simple. 
%Theorem on fiber bundle maps
\begin{theorem}
${\cal A}_{C\!\hbox{\footnotesize -flat}} \subset \bar{\cal A}_M$ is not left invariant by the pull back of fiber bundle maps. Instead 
\[
f^{\star}({\cal A}_{C\!\hbox{\footnotesize -flat}}) = 
{\cal A}_{\tilde{f}^{-1}C\!\hbox{\footnotesize -flat}}
\]
where $\tilde{f}$ is the 
map induced by 
$f$ in the base space. 
\end{theorem}
We leave the simple proof of this theorem to the reader. 

In this work the relevant fiber bundle maps are the ones that preserve the space of cellular decompositions of the base space that we are considering. Thus, we should focus either on piecewise analytic (also called stratified analytic) or on piecewise linear maps.  However, we loosely refer to them as ``diffeomorphisms."  

For convenience we write ${\cal A}_{C\!\hbox{\footnotesize -flat}}/\bar{\cal G}_{M,\star}$ as $\left( \frac{\cal A}{\cal G} \right)_C$. 

If the theory under study has diffeomorphism symmetry our construction 
of spaces of effective configurations 
breaks that symmetry. However, we can define a space of ``diffeomorphism" invariant configurations by injecting 
$\left( \frac{\cal A}{\cal G} \right)_C$ 
into 
$\bar{\cal A}_M/\bar{\cal G}_{M,\star}$ and using the notion of ``diffeomorphism" equivalence there. 
\begin{definition}
Two connections modulo gauge 
$[A_1], [A_2] \in \left( \frac{\cal A}{\cal G} \right)_C$ are said to be ``diffeomorphism" equivalent, 
\[
[A_1] \sim_{d} [A_2] , 
\]
if and only if there is a ``diffeomorphism" $\tilde{f}$ of $M$ 
(that has $\star \in M$ as a fixed point and)  
such that
$
\tilde{f}^{\star} [A_1] = [A_2]
$. 
(We are considering the action $\tilde{f}^{\star}$ defined by 
$f^{\star}: 
\bar{\cal A}_M / \bar{\cal G}_{M,\star} \to 
\bar{\cal A}_M / \bar{\cal G}_{M,\star}$
for any bundle map 
$f$ that induces $\tilde{f}$ on the base space and whose restriction to the fiber over 
$\star \in M$ is the identity.) 

Consider two cellular decompositions $C_1$ and $C_2 = \tilde{f}(C_1)$ (with 
$\tilde{f}(\star) = \star$). Clearly the quotient spaces 
$\frac{\left( \frac{\cal A}{\cal G} \right)_{C_1} }{ \sim_{d}}$ and 
$\frac{\left( \frac{\cal A}{\cal G} \right)_{C_2} }{ \sim_{d}}$ are identified by $\tilde{f}^{\star}$. Moreover, the identification would be the same for any other ``diffeomorphism" which relates the two cellular decompositions and has $\star \in M$ as a fixed point. 

Thus, we define 
the space of ``diffeomorphism" invariant effective configurations at scale $[C]$ as  
\[
\left( \frac{\cal A}{{\cal G}\times{\cal D}} \right)_{[C]} \doteq 
\frac{\left( \frac{\cal A}{\cal G} \right)_C }{ \sim_{d}} ,
\]
for any $C$ in the ``diffeomorphism" equivalence class $[C]$ (relative to the subgroup of ``diffeomorphisms" that fixes $\star \in M$). 
\end{definition}

It is important to remark that the equivalence relation $\sim_{d}$ is not induced by the space of automorphisms of the abstract cellular complex underlaying $C$. Even for a regular cellular decomposition whose automorphism group is large, $\sim_{d}$ is larger than the equivalence relation induced by automorphisms of $C$. 

There is a construction of effective configurations 
-- ${\cal A}_{K\!\hbox{\footnotesize -flat}} \subset \bar{\cal A}_M$ (where $K$ is an abstract cellular complex) --
which does not break the  ``diffeomorphism" symmetry. One 
can take quotient by the  ``diffeomorphism" group of 
${\cal A}_{K\!\hbox{\footnotesize -flat}}$
to find a space of ``diffeomorphism" invariant configurations finding the same space that we just defined \cite{K-flat}. 

From the point of view of the framework mentioned above, the framework 
presented in this article is {\em partially gauge fixed with respect to the ``diffeomorphism" group}. The complex nature of the equivalence relation $\sim_{d}$ in 
$\left( \frac{\cal A}{\cal G} \right)_C$ as compared to the transparent action of the ``diffeomorphism" group in $\bar{\cal A}_M$ 
means that this partial gauge fixation would not help if we want to analyze the ``diffeomorphism" symmetry. However, many other aspects (including those  presented in this paper) 
are much simpler to study in the gauge fixed framework. 

\section*{\large Acknowledgments}
We would like to thank  Elisa Manrique, Robert Oeckl and Axel Weber for 
the many hours of discussion about this subject. 
We also acknowledge exchange of ideas and constructive criticism from 
Michael Reisenberger, 
Abhay Ashtekar, 
Hanno Sahlmann, 
Danielle Perini, 
Fotini Markopoulou
and Lee Smolin 
during different stages of the development of this work. 
J. A. Z. thanks Mary-Ann for her invaluable support. 

This work was partially supported by grants CONACyT 40035-F, DGAPA IN108803-3 and  SNI 21286. 

\section*{Appendix}

For the convenience of the reader we 
give a minimal recollection of definitions and properties needed in the main body of the paper. The subjects covered are: 
\begin{enumerate}
\item Generalized connections
\item Cellular decompositions of a manifold and the limit $C \to M$
\end{enumerate}

%PRELIM OF GEN. CONNECTIONS AND CELL DECS. 
\subsubsection*{Generalized connections}
Generalized connections lie at the core of loop quantized theories. 

Consider a $G$-principal 
fiber bundle $(E, \varphi, M)$. 
In the physics literature it is customary to denote the space of smooth $G$-connections on that fiber bundle simply by ${\cal A}_M$. However, topologically different $G$-bundles over $M$ lead to different spaces of smooth connections. 

A {\em generalized connection} $A\in \bar{\cal A}_M$ is an assignment 
(a semigroup morphism) 
of holonomies to oriented paths in $M$ 
\[
A(\gamma) : \varphi^{-1} (s(\gamma)) \to \varphi^{-1} (t(\gamma)) , 
\]
where $s(\gamma)$ and $t(\gamma)$ denote the source and target points of $\gamma$. 
A holonomy must be compatible with the right $G$-action on the fiber bundle, 
$A(\gamma) [x g] = A(\gamma) [x] g$ for any $g \in G$ and any 
$x \in \varphi^{-1} (s(\gamma))$. 
Holonomies can be composed when the corresponding paths can. Asking that the assignment be a morphism means that $A(\gamma_2 \circ \gamma_1) = A(\gamma_2) \circ A(\gamma_1)$. 
We remark that paths of the form $c^{-1} \circ c$ are regarded as equivalent to the path that stays at the  single point $s(c)$; thus $A(c^{-1}) = A(c)^{-1}$. 

It is important to emphasize that there are no requirements of smoothness or even continuity 
for the holonomy assignments with respect to deformations of the path. 
Analogously the group of gauge transformations acts on holonomy assignments without any continuity restrictions. 
If one imposes restrictions of this type on the space of connections and the space of gauge transformations, one can recover the space of smooth connections modulo gauge. 

At the gauge invariant level the focus changes to group homomorphisms from the group of $\star \in M$-based oriented loops (modulo an equivalence relation that prevents holonomical triviality), 
${\cal P}_{M, \star}$, 
to $G$. The space of such homomorphisms is the space of gauge equivalence classes of generalized connections 
\[
\bar{\cal A}_M / \bar{\cal G}_{M,\star}  = {\rm Hom}({\cal P}_{M, \star}, G) .
\]
Again no continuity restrictions are imposed. If one imposes this type of conditions one can reconstruct the space of smooth connections modulo gauge. In fact, 
it is in exactly this terms that the reconstruction theorems are enounced and proven  \cite{reconstruction}. 

Instead of placing continuity restrictions from classical considerations, one defines a different topology for $\bar{\cal A}_M / \bar{\cal G}_{M,\star}$. By definition this is the weakest topology for which 
cylindrical functions, ${\rm Cyl}(\bar{\cal A}_M / \bar{\cal G}_{M,\star})$, are continuous. Cylindrical functions are functions on 
$\bar{\cal A}_M / \bar{\cal G}_{M,\star}$ induced by continuous functions on finitely many copies of $G$ by the choice of a based loop for each copy of $G$. 
With respect to this topology, the space $\bar{\cal A}_M / \bar{\cal G}_{M,\star}$ is independent of the topology of the total space of the fiber bundle. 
This finishes with our definition of 
$\bar{\cal A}_M / \bar{\cal G}_{M,\star} $.

\subsubsection*{Cellular decompositions and the limit $C \to M$}

A typical example of a cellular decomposition is a triangulation of the sphere whose cells are: four triangles, six edges and four vertices. 

A {\em cellular decomposition} $C$ of a manifold $M$ 
is a presentation of it as a union of disjoint cells 
\[
M= \cup_{c_{\alpha \in C}} c_{\alpha} , 
\]
\[
 c_{\alpha} \cup  c_{\beta} = \emptyset \hbox{ if } \alpha \neq \beta .
\]
Each cell $c_{\alpha}$ is the image of 
an open convex polyhedron in $\R^{n({\alpha})}$ with $n({\alpha})$ between 
zero and $\dim M$. Then, we can specify a cell by a pair consisting of an open convex polyhedron in $\R^n$ and a map that takes the polyhedron to $M$, 
\[
c_{\alpha} = ( p_{\alpha}, \phi_{\alpha} ) . 
\]
To be precise we have to say what kind of maps do we consider. 
In this work we consider maps $\phi_{\alpha} : \R^{n({\alpha})} \to M$ which are either piecewise analytic or piecewise linear. 

The $n$-skeleton of a cellular decomposition $C^{(n)}$ is the collection of cells of dimension smaller or equal to $n$. For example, $C^{({\rm dim} M)}$ is $C$ and $C^{(0)}$ is the set of vertices of $C$. In our work we use the one dimensional complex $C^{(1)}$ because it is a lattice. 

Notice that the set of cellular decompositions of a manifold admits a partial order relation that tell us if a cellular decomposition $C_2$ is finer than another one $C_1$. We write 
\[
C_1 \leq C_2
\]
if any cell in the coarser decomposition is a finite union of cells of the finer decomposition. 

Triangulations (or simplicial decompositions) are particular examples of cellular decompositions where the polyhedra used are only simplices. In this category of cellular decompositions one can easily define the {\em barycentric subdivision} operation that produces a finer cellular decomposition, 
$\Delta \leq {\rm Sd}(\Delta)$. The new triangulation ${\rm Sd}(\Delta)$ is easily 
constructed inductively: 
\begin{enumerate}
\item 
To each simplicial cell 
$\delta_{\alpha} \in \Delta$ one assigns a zero dimensional cell in ${\rm Sd}(\Delta)$
called the barycenter of $\delta_{\alpha}$. (For zero dimensional cells the barycenter coincides with the original cell.) 
\item 
For one dimensional simplicial cells the cone over their boundary with vertex in their barycenter is a simplicial subdivision of $\delta_{\alpha}$. We call it the 
barycentric subdivision of $\delta_{\alpha}$, 
${\rm Sd}(\delta_{\alpha})$. 
\item 
The barycentric subdivision of the union of cells is by definition the union of the barycentric subdivision of the cells, ${\rm Sd}(\delta_{\alpha}\cup \delta_{\beta}) = 
{\rm Sd}(\delta_{\alpha}) \cup {\rm Sd}(\delta_{\beta})$. 
\item 
Assume that ${\rm Sd}$ is defined for cells of dimension $n-1$. 
\item 
The boundary of an 
$n$ dimensional simplicial cell $\partial (\delta_{\alpha})$
is the union of $n-1$ dimensional simplicial cells. 
The cone over \\
${\rm Sd}(\partial (\delta_{\alpha}))$ with vertex in the barycenter of 
$\delta_{\alpha}$ is defined as ${\rm Sd}(\delta_{\alpha})$. 
\end{enumerate}

%%%%%%
The partial order of cellular decompositions is {\em directed} in the direction of refinement. This means that given any two cellular decompositions $C_1, C_2$ there is third one such that $C_1 \leq C_3$ and $C_2 \leq C_3$. 
With this property in mind is that we restrict to piecewise analytic or piecewise linear cellular decompositions; the directionality property would not hold if we allowed any smooth map. 

The directionality property is the one that will let us talk about the continuum limit as a limit towards ``the finest cellular decomposition." For example if there is any physically meaningful number calculated in an effective theory associated to a cellular decomposition $N(C)$, we would like it to have a finite limit when we remove the cut-off. Our notation is 
$N(M) = \lim_{C\to M} N(C)$, where the directionality property of the partial order gives a meaning to the limit ``in terms of epsilons and deltas as in ordinary calculus." When one is talking about objects different than numbers one has to specify the meaning of $\lim_{C\to M}$; here we simply remark that the directionality of the partial order makes it possible to define a variety of such limits.

%\section{References}

\end{document}